\documentclass[12pt]{article}

\newcommand{\beq}{\begin{equation}}
\newcommand{\eeq}{\end{equation}}
\newcommand{\bea}{\begin{eqnarray}}
\newcommand{\eea}{\end{eqnarray}}
\newcommand{\half}{{\scriptstyle{{1\over 2}}}}
\newcommand{\tr}{\mbox{\,tr\,}}
\newcommand{\ad}{{\rm ad}}
\newcommand{\norm}[1]{\left\| #1 \right\|}

\newcommand{\Lm}{\Lambda}

\newcommand{\Ss}[1]{\mbox{$\cal #1$}}
\newcommand{\pr}{\partial}

\newcommand{\Order}[1]{\Ss{O}\left(#1\right)}

\newcommand{\Ref}[1]{(\ref{#1})}
\def\cD{{\cal{D}}}

\begin{document}

\hfill\hbox{MPI-PhT/02-60, INLO-PUB-06/02}
\vskip1cm
\begin{center}{\Large\bf Defining $\langle A^2\rangle$ in the
finite volume hamiltonian formalism}\\[1cm]
{\bf L. Stodolsky${}^1$, Pierre van Baal${{}^2}$ and V.I.
Zakharov${{}^1}$}\\[3mm]
${}^{1)}${\em Max-Planck Institut f\"ur Physik, F\"ohringer Ring 6,
D-60805 M\"unchen, Germany}\\
${}^{2)}${\em Instituut-Lorentz for Theoretical Physics, University
of Leiden, P.O.Box 9506,\\ NL-2300 RA Leiden, The Netherlands}
\end{center}

\section*{Abstract}
It is shown how in principle for non-abelian gauge theories it is
possible in the finite volume hamiltonian framework to make sense 
of calculating the expectation value of $\norm{A}^2=\int d^3x\left(
A^a_i(\vec x)\right)^2$. Gauge invariance requires one to replace 
$\norm{A}^2$ by its minimum over the gauge orbit, which makes it a 
highly non-local quantity. We comment on the difficulty of finding 
an expression for $\norm{A}^2_{\rm min}$ analogous to that found 
for the abelian case, and the relation of this question to  Gribov 
copies. We deal with these issues by implementing the hamiltonian 
on the so-called fundamental domain, with appropriate boundary 
conditions in field space, essential to correctly represent the 
physics of the problem. 

\section{Introduction}
The question of a dimension-two condensate in QCD or in pure 
gauge theories~\cite{Cond}, is a fascinating one. Condensates 
are believed to be connected to the non-perturbative structure 
of the theory. In a theory without manifest dimensional parameters 
the only evident operator for such a condensate is in terms of the 
vector potential itself $V^{-1}\norm{A}^2$. This is certainly the 
simplest object one can imagine in a pure gauge theory, but it
suffers from the obvious defect that it is not gauge invaraint 
and so cannot have any physical meaning. In perturbation theory 
this can be remedied when one replaces the vector potential by 
its transverse part, satisfying $\partial_\mu A^\mu(x)=0$.

Irrespective of the question if one could separate off the 
perturbative contribution, it is not at all clear if one can even 
make sense of calculating $\langle A^2\rangle$ non-perturbatively.
It is only the latter point we address in this paper, by using the 
hamiltonian framework. An essential ingredient is to consider 
$\norm{A}^2_{\rm min}$, which minimizes $\norm{A}^2$ along the 
gauge orbit. In the abelian case this can be expressed in terms 
of the field strength~\cite{GSZ}, with some non-locality involved. 
In Sect.~2 we point out that the difficulty to find a similar
formula for $\norm{A}^2_{\rm min}$ in the non-abelian case is 
related to the Gribov copy problem~\cite{Gri}. 

We will then show in Sects.~3 and 4 how the hamiltonian formulation in a
finite volume, using the Coulomb gauge~\cite{Chle}, can deal with the 
Gribov issue by restricting the transverse gauge fields to the fundamental 
domain~\cite{Sefr,Del1}. On its boundary, $\norm{A}^2_{\rm min}$ is 
degenerate~\cite{Vba}, and it is this that allows one to consistently 
define $\langle A^2\rangle=\langle 0|V^{-1}\norm{A}^2_{\rm min}|0\rangle$.
Non-perturbative contributions typically arise when the wave functional 
starts to reach the boundary of the fundamental domain. This has been 
successfully implemented in the past to calculate the low-lying spectrum 
in a finite volume (for a review see Ref.~\cite{Iof}).

The finite volume cutoff allows us to define the contribution coming 
from the low-lying modes for which the wave functional is affected by 
the boundary of the fundamental domain, in terms of an effective 
hamiltonian for these modes. By excluding the perturbative contributions
coming from the modes that are integrated out, in as far as they do
{\em not interact} with the low-lying modes kept in the effective 
hamiltonian, this gives by construction a finite result. This is 
presented in Sect.~5, and illustrates that $\langle A^2\rangle$ 
indeed receives a non-perturbative contribution due to the boundary 
conditions in field space, properly reflecting the non-trivial geometry 
of the configuration space~\cite{Bavi}. 

We resist the temptation of subtracting the perturbative result 
obtained with the effective hamiltonian, to avoid the usual difficulties 
with defining a condensate unambiguously. The aim of this paper is to 
demonstrate that there is a way to define $\langle A^2\rangle$ beyond 
perturbation theory. 

\section{Abelian gauge theories}
An elegant expression for abelian gauge theories exists, that
splits $\norm{A}^2$ in transverse and longitudinal parts, which 
when expressed in momentum space (the Fourier components are 
denoted by a tilde)
\beq
\norm{A}^2=\int d^np~\frac{\tilde F^{\mu\nu}(p)\tilde
F^*_{\mu\nu}(p)}{p_\mu p^\mu}+\int d^np~\frac{p^\mu
\tilde A_\mu(p)p^\nu\tilde A^*_\nu(p)}{p_\mu p^\mu},
\eeq
holds in any dimension~\cite{GSZ}. It relies on the well-known
vector identity
\beq
\half\tilde F^{\mu\nu}(p)\tilde F^*_{\mu\nu}(p)=p_\mu p^\mu\tilde A^\nu(p)
\tilde A^*_\nu(p)-p^\mu p^\nu \tilde A_\mu(p) \tilde A^*_\nu(p),
\eeq
useful in setting up perturbation theory. Since minimizing
$\norm{A}^2$
along the gauge orbit implies the gauge field at this minimum
satisfies the gauge condition $\partial^\mu A_\mu(x)=0$, or 
$p^\mu\tilde A_\mu(p)=0$, this implies we have a gauge invariant 
expression for $\norm{A}_{\rm min}^2$,
\beq
\norm{A}_{\rm min}^2=\int d^np~\frac{\tilde F^{\,2}_{\mu\nu}(p)}{p_\mu p^\mu},
\eeq
with only a limited amount of non-locality.

There is a small problem to address here because $p^2$ cannot be
inverted for zero-momentum. In coordinate space this could give
rise to boundary terms~\cite{GSZ}. The problem persists with 
periodic boundary conditions also. This is interesting in that it 
reveals a subtle issue related to Gribov copies~\cite{Gri}, which 
are essential to the non-abelian problem.

With boundary conditions periodic in a length $L$, the integral over 
momenta is replaced by a sum. The zero-momentum components of 
$F_{\mu\nu}$ and $\partial_\mu A^\mu$ vanish, but not those for 
$A_\mu$. This means that we have to replace $\norm{A}^2_{\rm min}$ 
by $\norm{A}^2_{\rm min}-\tilde A^*_\mu(0) \tilde A^\mu(0)= 
\norm{A}^2_{\rm min}-L^{-n}(\int d^nx A_\mu(x))^2$.  This may seem 
an insignificant modification as $L\to\infty$, but it is exactly 
what is needed to deal with the problem of Gribov copies:  different 
gauge fields related by a gauge transformation that satisfy the 
gauge condition $\partial_\mu A^\mu(x)=0$, but for which the value 
of $\norm{A}^2$ differ. 

These Gribov copies can even be present in the abelian theory (for 
finite volume and periodic boundary conditions). However, the 
difference in $A$ between Gribov copies in this case has zero 
momentum. After all, with $[h]A_\mu(x)$ the gauge field obtained 
by a gauge transformation $h(x)$ from $A_\mu(x)$, requiring 
$\partial_\mu ([h]A^\mu(x))=\partial_\mu A^\mu(x)=0$ implies 
$\partial^2_\mu\log h(x)=0$. This fixes the allowed gauge 
transformations to be of the form $h(x)=\exp(2\pi ix^\mu n_\mu/L)$, 
with $n_\mu$ integer (as imposed by the periodic boundary conditions). 
It means that the difference in $\norm{A}^2$ between different Gribov 
copies is only in the zero-momentum component and we have
\beq
\norm{A}^2-L^{-n}(\int d^nx A_\mu(x))^2
=\sum_{p\neq0}\frac{\tilde F^{\mu\nu}(p)\tilde F^*_{\mu\nu}(p)}{p_\mu
p^\mu} +\sum_{p\neq0}~\frac{p^\mu\tilde A_\mu(p)p^\nu\tilde 
A^*_\nu(p)}{p_\mu p^\mu},
\eeq
and
\beq
\norm{A}_{\rm min}^2-L^{-n}(\int d^nx A_\mu(x))^2=\sum_{p\neq0}
\frac{\tilde F^{\mu\nu}(p)\tilde F^*_{\mu\nu}(p)}{p_\mu p^\mu}.
\eeq
This formula is also correct when instead of taking the absolute
minimum we have a stationary point. It is this that causes the 
Gribov problem: there are many stationary points of $\norm{A}^2$ 
along a given gauge orbit, where at each of these stationary points 
the (Coulomb or Landau) gauge condition $\partial_\mu A^\mu(x)=0$
holds. For abelian gauge theories, this problem only occurs in 
a finite volume with periodic boundary conditions and the Gribov 
copies can be fully classified in terms of the zero-momentum 
component of the gauge field.

It is amusing to observe, since the vector identity is still true,
that in the non-abelian theory (in a finite volume with periodic 
boundary conditions) we may again write
\beq
\norm{A}^2-L^{-n}(\int d^nx A^a_\mu(x))^2 =\sum_{p\neq0}\frac{\tilde 
f^{a\mu\nu}(p)\tilde f^{a*}_{\mu\nu}(p)}{p_\mu p^\mu}+\sum_{p\neq0}~
\frac{p^\mu\tilde A^a_\mu(p)p^\nu\tilde A^{a*}_\nu(p)}{p_\mu p^\mu},
\eeq
with $f^a_{\mu\nu}=\partial_\mu A^a_\nu(x)-\partial_\nu A^a_\mu(x)$
and
\beq
\norm{A}_{\rm min}^2-L^{-n}(\int d^nx A^a_\mu(x))^2=\sum_{p\neq 0}
\frac{\tilde f^{a\mu\nu}(p)\tilde f^{a*}_{\mu\nu}(p)}{p_\mu p^\mu}.
\eeq
This reveals at once the problem for the non-abelian case: the
right hand side involving only the ``curl" part of the field 
tensor is not gauge invariant, and the problem of Gribov copies 
cannot be restricted to the zero-momentum component of the gauge 
field. Indeed explicit examples are known that illustrate this 
point~\cite{Vba}. Minimizing along the gauge orbit has the 
complexity of a spin glass problem, with many local minima,
which from the computational point of view makes it in practice
impossible to identify the absolute minimum.

\section{Non-abelian gauge theories}

We have seen that there appears to be no simple  gauge invariant
expression for $\norm{A}^2_{\rm min}$ in non-abelian gauge
theories, even allowing for non-locality. Certainly a formula 
for $\norm{A}^2$ similar to the abelian case, where we have a 
gauge invariant expression plus something vanishing at the 
stationary points of $\norm{A}^2$ cannot apply. This would give 
the same value for $\norm{A}^2$ at Gribov copies, while we know 
on the contrary that generically $\norm{A}^2$ is different for 
such copies. 

Therefore we would like to turn to the hamiltonian picture of 
non-abelian gauge theory to provide some insight into the 
question of $\norm{A}^2_{\rm min}$. What we would like to 
find is that in a certain sense  $\norm{A}^2_{\rm min}$ viewed 
as a quantum-mechanical operator can have a non-trivial 
expectation value, as we shall now explain.

In the  hamiltonian picture~\cite{Chle}, where $A_0=0$, one considers 
wave functionals on field space. The ``coordinates" are the spatial 
components of the vector potential at every point in ordinary space 
$\vec x$, $A_i^a(\vec x)$. In a lattice formulation, for example, 
there can be a finite number of coordinates. Or in momentum space 
one may use $\tilde A_i^a(\vec k)$ as the variables. In any event 
we imagine a wave function in these coordinates $\Psi(A)$.

Now in the simplest case of  a free abelian theory, one has a problem 
equivalent to the ordinary harmonic oscillator for the modes in 
momentum space. Evidently, there is a non-zero value of $\norm{A}^2$ 
for each mode, since this is just the spread of the wave function in 
the ground state of the oscillator. This is what one expects for the 
ordinary vacuum and we shall call this the ``perturbative contribution'', 
since when summed over all modes it represents the well known (infinite)  
zero-point motion of free fields.

When $A$ becomes large $\Psi$ will be sensitive to the $A^3$ and 
$A^4$ terms in the potential energy $\half B_i^a(\vec x)^2$, with 
$B_i^a(\vec x)$ the Yang-Mills magnetic field. These correspond to 
the non-linearities of the theory and determine the directions in 
which $\Psi(A)$ can spread. When $\Psi(A)$ can no longer be 
neglected near the boundary of the fundamental domain, it  is 
sensitive to the boundary conditions required to make the problem 
well defined. It is this that can no longer be described in 
perturbation theory, leading to non-perturbative contributions. 

The implementation of the hamiltonian approach depends very much on 
whether and how the gauge condition $\partial_i A^i(\vec x)=0$ is
handled. In principle, one may not apply any condition of this type
at all and simply assume that the $\Psi$ is constant in gauge
directions, that is, constant over a gauge orbit. Although this is
conceptually simple and is usually the approach in lattice
simulations, it is remote from standard perturbation theory and
difficult to apply for concrete hamiltonian calculations. In
particular for our present question this would necessitate an
explicit gauge invariant expression  for an $\norm{A}^2_{\rm min}$
operator, which as explained in the previous section, we do not have.

The more common approach in the hamiltonian method is thus to
formulate the problem in terms of one representative $A$ field
configuration on a gauge orbit in order to reduce the number of 
variables. This configuration is found by imposing $\partial_i 
A^i(\vec x)=0$ and one then uses the Faddeev-Popov method to find 
the volume of the gauge orbit when integrating over $A$ 
configurations~\cite{Bavi}. However, this leads to the problem of 
Gribov copies, since in the non-abelian case there is more than 
one $A$ configuration with $\partial_i A^i(\vec x)=0$ on a given 
gauge orbit. For the question of $\norm{A}^2_{\rm min}$ the 
existence of Gribov copies means that $\partial_i A^i(\vec x)=0$ 
no longer determines that configuration where $\norm{A}^2$ is an 
absolute minimum.

A way of dealing with these complications is to define a fundamental 
domain~\cite{Sefr} where there is only one $A$ configuration with 
$\partial_i A^i(\vec x)=0$ on each gauge orbit. Restricting the 
variables to this domain leads to a well-defined quantum-mechanical 
problem, where we may calculate the expectation value of $A^2$. 
The price of this simplification, however, is a complicated topology 
in $A$ space on the boundary of this domain, as we shall now explain. 
To keep things well defined, we introduce a finite volume in ordinary 
space as an infrared cutoff, like the torus $T^3$ or the sphere $S^3$. 
For the torus, zero-momentum modes have to be treated carefully, but one 
has learned how to deal with this, see a recent review in Ref.~\cite{Iof}.

The hamiltonian formalism provides more intuition on how to deal with 
non-perturbative contributions in situations where semi-classical 
techniques can no longer be used. The high energy modes can be 
well-approximated by harmonic oscillator contributions to the wave 
functional. In the direction of these field modes the potential energy
rises steeply. Their contributions, which include regulating the
ultraviolet behavior, can presumably be treated perturbatively, in
particular giving rise to the running of the coupling constant.

The finite volume allows us to have a well-defined mode expansion
in momentum space. Due to the classical scale invariance, the 
hamiltonian can be formulated in terms of dimensionless fields. 
This can be extended to the quantum theory, as Ward identities 
allow for a field definition without anomalous scaling. Thus we 
absorb the bare coupling constant in the gauge field. In these 
conventions the field strength is given, in terms $A_i(\vec x)=
A^a_i(\vec x)T^a$ ($T^a$ the anti-hermitian generators, 
normalized according to $\tr(T^aT^b)=-\half\delta_{ab}$), by
\beq
F_{ij}(\vec x)=F^a_{ij}(\vec x)T^a=\pr_i A_j(\vec x)-\pr_j 
A_i(\vec x)+[A_i(\vec x),A_j(\vec x)]\label{eq:F}
\eeq
and the hamiltonian density reads 
\beq
{\cal H}(\vec x)=-\half g^2\left(\frac{\partial}{\partial A^a_j
(\vec x)}\right)^2+\frac{1}{2g^2}\left(B^a_j(\vec x)\right)^2,
\label{eq:ham}
\eeq
where $B^a_k(\vec x)=\half\varepsilon_{ijk}F^a_{ij}(\vec x)$. 
When all fields and coordinates are expressed in units of $L$
(with $Q_i^a(\vec x)$ the usual expression for the gauge field,
$A_i^a(\vec x)=gLQ_i^a(L\vec x)$), apart from the overall scaling 
dimension of the hamiltonian ($1/L$), only the running coupling 
introduces a non-trivial volume dependence~\cite{Iof}.

Therefore, the only sensitivity to the length scale $L$ is through an 
increasing coupling as we increase $L$. An increasing coupling will 
cause spreading of the wave functional, simply because the overall 
strength of the potential (proportional to $1/g^2$) is reduced. The 
essential additional ingredient required to address non-perturbative 
effects is the boundary conditions in {\em field space}, at the 
boundary of the fundamental domain. Only in this way can gauge 
invariance be implemented properly at all stages. On the other hand,
asymptotic freedom guarantees that in small volumes the running 
coupling is small and it thus keeps the wave functional localized 
near the classical vacuum manifold. What has become clear~\cite{Iof} 
is that the transition from finite to infinite volume is driven by 
field fluctuations that cross the barrier which is associated with 
tunneling between different classical vacua. This is natural, since 
this barrier (the finite volume sphaleron, which will typically lie 
on the boundary of the fundamental domain), will be the direction 
beyond which the wave functional can first spread most significantly, 
as it provides the lowest mountain pass in the energy landscape.

\section{The fundamental Domain}

We will summarize how to completely fix the gauge and show that the
boundary of the fundamental domain, unlike its interior, has gauge 
copies that implement the non-trivial topology of field space. The 
essential observation that allows one to define $\norm{A}^2_{\rm 
min}$ as a proper gauge invariant quantity is that this minimum along 
the gauge orbit is degenerate when the associated gauge fields (by
definition related by a gauge transformation) represent points on the 
boundary of the fundamental domain that are to be identified.

Restricting to three space dimensions, we will now be a bit more
precise about how to minimize the $L^2$ norm of the vector potential 
along the gauge orbit~\cite{Sefr,Del1} (recall that the vector 
potential $A_i(\vec x)$ is anti-hermitian).
\beq
\norm{[h]A}^2=-\int_M\tr\left(\left( h^{-1}(\vec x)
A_i(\vec x)h(\vec x)+h^{-1}(\vec x)\pr_i h(\vec x)\right)^2\right).
\label{eq:gAnorm}
\eeq
The integral over the finite spatial volume $M$ is with the 
appropriate canonical volume form. We introduce the short-hand 
notation $[h]A$ for a gauge transformation $h(\vec x)$. Expanding 
around the minimum of Eq.~\Ref{eq:gAnorm}, writing $h(\vec x)= 
\exp(X(\vec x))$ ($X(\vec x)$ is, like the gauge field 
$A_i(\vec x)$, an element of the Lie-algebra) one easily finds
\bea
\norm{\,[h]A}^2 &=& \norm{A}^2+2\int_M \tr(X \partial_i A_i)+\int_M
\tr(X^\dagger FP (A) X) \\&&\hskip-1.3cm+\frac{1}{3}\int_M\tr\left(
X\left[[A_i,X],\partial_i X\right]\right)+\frac{1}{12}\int_M\tr
\left([\Ss{D}_iX,X][\partial_i X,X]\right)+\Order{X^5},\nonumber
\label{eq:Xexp}
\eea
where $FP(A)$ is the Faddeev-Popov operator $(\ad(A)X\equiv[A,X])$
\beq
FP (A)=-\partial_i \Ss{D}_i(A)\equiv-\partial_i(\partial_i+
\ad(A_i)).\label{eq:FP}
\eeq

At a local minimum the vector potential is therefore transverse,
$\partial_i A_i=0$, and $FP(A)$ must be a positive operator. The 
set of all these vector potentials is by definition the Gribov 
region $\Omega$. Using the fact that $FP(A)$ is linear in $A$, 
$\Omega$ is seen to be a convex subspace of the set of transverse 
gauge fields $\Gamma$. Its boundary $\partial \Omega$ is called 
the Gribov horizon. At the Gribov horizon, the {\em lowest}
non-trivial eigenvalue of the Faddeev-Popov operator vanishes, 
and points on $\partial\Omega$ are associated with coordinate 
singularities. Any point on $\partial\Omega$ can be seen to have 
a finite distance to the origin of field space and in some cases 
even uniform bounds can be derived~\cite{Dezw,Zwan}.

The Gribov region is the set of {\em local} minima of the norm
functional, Eq.~\Ref{eq:gAnorm}, and needs to be further 
restricted to the {\em absolute} minima to form the fundamental 
domain~\cite{Sefr}, which will be denoted by $\Lambda$. The 
fundamental domain is clearly contained within the Gribov
region. To show that also $\Lambda$ is convex, we define an 
operator $FP_f(A)$ via
\bea
&&\norm{\,[h]A}^2-\norm{A}^2=\int_M\tr\left(A_i^2\right)-\int_M\tr
\left(\left(h^{-1}A_ih+h^{-1}\pr_ih\right)^2\right)\label{eq:FPf}
\label{eq:FDmin}\\&&\qquad\qquad=\int_M\tr\left(h^{-1}FP_f(A)\,h
\right),\quad FP_f(A)\equiv-\pr_i(\pr_i+A_i),\nonumber
\eea
(remember that in our conventions $\tr(A_i^2)$ is negative), where 
$FP_f(A)$ acts on Lie-group valued functions and is similar to the 
Faddeev-Popov operator (which acts on Lie-algebra valued functions). 
Both $FP(A)$ and $FP_f(A)$ are hermitian operators when $A$ is a 
stationary point of the norm functional, i.e. for $A$ transverse. 
The fundamental domain $\Lm$ is the set of gauge fields $A$ for which
Eq.~\Ref{eq:FDmin} has its minimum at zero when varying $h$ over the
gauge group. (If this minimum is unique it occurs for $h=1$.)
Using that $FP_f(A)$ is linear in $A$, the convexity of $\Lm$ is
automatic: A line connecting two points in $\Lm$ lies within $\Lm$.

If we would not specify anything further, since a convex space is
contractible, the fundamental region could never reproduce the 
non-trivial topology of the field space. This means that $\Lm$ 
should have a boundary~\cite{Vba}. Indeed, as $\Lambda$ is 
contained in $\Omega$, this means $\Lm$ is also bounded in each 
direction. Consider a gauge orbit and two gauge configurations on 
it, giving the absolute and first relative minimum of $\norm{A}^2$ 
respectively. In general the two configurations are connected 
by a finite, or even a ``big", gauge transformation. Now take
the ``ray" that extends from the relative minimum configuration 
to $A=0$. There it will have $\norm{A}^2=0$. Its gauge copy, 
initially an absolute minimum, will also vary continuously as we
go along the ray towards $A=0$, but will not in general have 
$\norm{A}^2=0$ at the end of the variation. Therefore the norms 
of each of these two configurations must pass each other during 
this variation. At the crossing we have degenerate minima of 
$\norm{A}^2$ at distinct points on the gauge orbit. These 
correspond to different points of the boundary of $\Lm$, 
identified by gauge equivalence. This gives the problem its 
non-trivial topology.

When $L$ denotes the linear size of the spatial volume $M$, we may 
express the gauge fields in the dimensionless combination of $L A$ 
(in our conventions the fields have no anomalous scale dependence), 
and {\em the shape and geometry of the Gribov and fundamental 
regions are scale independent}. We should note that the norm 
functional is degenerate along the constant gauge transformations 
and indeed the Coulomb gauge does not fix these gauge degrees of 
freedom. We simply demand that the wave functional is in the singlet 
representation under the constant gauge transformations. Therefore, 
with $G$ the gauge group, $\Lm/G$ represents the gauge invariant 
configuration space, for which $\Lm$ is assumed to include the 
non-trivial boundary identifications that restore the non-trivial 
topology of this space.

\begin{figure}[hb]
\vspace{5.6cm}
\includegraphics{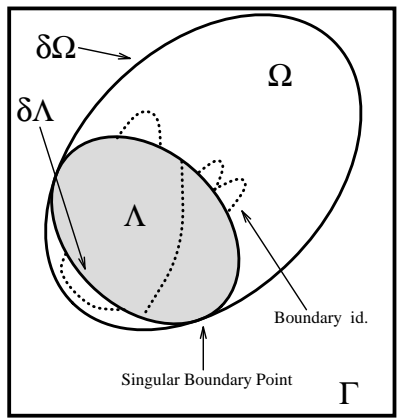}
\caption{Sketch of the fundamental (shaded) and Gribov regions, 
embedded in the space of transverse gauge fields ($\Gamma$). 
The dotted lines indicate boundary identifications.}
\label{fig:Grib}
\end{figure}

If a degeneracy at the boundary is continuous, other than by
constant gauge transformations, one necessarily has at least one 
non-trivial zero eigenvalue for $FP(A)$ and the Gribov horizon 
will touch the boundary of the fundamental domain at these 
so-called singular boundary points. We sketch the general
situation in Fig.~\ref{fig:Grib}. In principle, by choosing a
different gauge fixing in the neighborhood of these points one 
could resolve the singularity. If singular boundary points would 
not exist, all that would have been required is to complement the 
hamiltonian in the Coulomb gauge with the appropriate boundary 
conditions in field space. Since the boundary identifications are
by gauge transformations the boundary condition on the wave
functionals is simply that they are identical under the boundary 
identifications, possibly up to a phase in case the gauge 
transformation is homotopically non-trivial.

Singular boundary points are to be expected~\cite{Vba}. Generically, 
at singular boundary points the norm functional undergoes a 
bifurcation moving from inside to outside the fundamental (and Gribov) 
region. The absolute minimum turns into a saddle point and two local 
minima appear. These are necessarily gauge copies of each other. The 
gauge transformation is homotopically trivial as it reduces to the 
identity at the bifurcation point, evolving continuously from there on.

The necessity to restrict to the fundamental domain, a subset of
the transverse gauge fields, introduces a non-local procedure in 
field space. This cannot be avoided since it reflects the 
non-trivial topology of this space. We stress again that its 
topology and geometry are scale independent. Homotopical
non-trivial gauge transformations are in one to one correspondence
with non-contractible loops in field space, which give rise to 
conserved quantum numbers. The quantum numbers are like the Bloch 
momenta in a periodic potential and label representations of the 
homotopy group of gauge transformations. On the fundamental domain 
the non-contractible loops arise through identifications of 
boundary points. Although slightly more hidden, the fundamental 
domain will therefore contain all the information relevant for the 
topological quantum numbers. Sufficient knowledge of the boundary
identifications will allow for an efficient and natural projection
on the various superselection sectors. Typically we integrate out 
the high-energy modes, being left with the low-energy modes whose 
dynamics is determined by an effective hamiltonian defined on the 
fundamental domain (restricted to these low-energy modes). In this 
it is assumed that the contributions of the high-energy modes can 
be dealt with perturbatively, generating the running coupling and 
the effective interactions of the low-energy modes.

With the boundary identifications implemented, and the fact that by
construction $\norm{A}_{\rm min}^2$ respects these boundary
identifications,
\beq
\langle0|\norm{A}_{\rm min}^2|0\rangle\equiv\int_\Lm\mu(A)\cD A~
\Psi_0^*(A)\norm{A}_{\rm min}^2\Psi_0(A)
\eeq
is in principle well-defined, and could form the basis for
establishing the existence of a non-perturbative dimension two 
condensate. Here $\Psi_0(A)$ is the groundstate wave functional, 
and $\mu(A)$ is the appropriate measure on field 
space~\cite{Chle,Bavi}, the integral assumed to be confined to 
the fundamental domain $\Lm$ of the transverse gauge fields.

\section{Small volume results}

In a small volume with periodic boundary conditions the running
coupling is small and one can use perturbation theory. To lowest 
order the wave functional is simply a product of harmonic oscillators 
for each of the field modes. The zero-momentum modes, however,
need to be treated separately since the potential in this sector
is quartic. In computing $\langle0|\norm{A}_{\rm min}^2|0\rangle$ 
we use for these zero-momentum modes the groundstate wave function 
belonging to the hamiltonian of Eq.~\Ref{eq:ham}, truncated to the 
zero-momentum modes, 
\beq
H_0=-\frac{g^2}{2L}\left(\frac{\partial}{\partial c^a_j}\right)^2
-\frac{1}{2g^2L}\tr\left([c_j,c_k]^2\right)
\eeq
where we defined $L^{-3}\int d^3x A_j^a(\vec x)=c_j^a/L$ and 
$c_j=c_j^aT^a$. A simple rescaling of the fields with a factor 
$g^{2/3}$ shows that the energies of this truncated hamiltonian 
are proportional to $g^{2/3}/L$ and that the zero-momentum 
contribution to $\langle A^2\rangle\equiv L^{-3}\langle0|
\norm{A}_{\rm min}^2|0\rangle$ is proportional to $g^{4/3}/L^2$.

It is in the direction of the zero-momentum modes that the wave functional 
will first reach the boundary of the fundamental domain for increasing 
coupling (due to an increase in the volume). There is no classical 
potential along the direction of the abelian zero-momentum modes (for 
which the commutator $[c_i,c_j]$ vanishes) and the boundary of the 
fundamental domain in these abelian zero-momentum components can be 
shown~\cite{Iof} to occur at $(c_i^a)^2=\pi^2$. L\"uscher~\cite{Lue} 
has derived an effective hamiltonian for the zero-momentum modes that 
incorporates the higher order corrections due to the interactions with 
the non-zero momentum modes, using so-called Bloch perturbation 
theory~\cite{Bloch}. Amongst other things, this turns the bare coupling 
constant into a running coupling $g(L)$. It is beyond the scope of this 
paper to describe the details of this calculation, but we remark that an 
efficient way to compute $\langle0|\norm{A}_{\rm min}^2|0\rangle$ is 
by adding $\lambda\left(A_j^a(\vec x)\right)^2$ to the hamiltonian 
density of Eq.~\Ref{eq:ham}. Here $A_j^a(\vec x)$ is assumed to be 
transverse and to lie within the fundamental domain. We will consider 
volumes where the restriction to the fundamental domain is felt for the 
zero-momentum modes only, such that we can integrate out the non-zero 
momentum modes perturbatively. The resulting effective hamiltonian 
will contain a term depending on $\lambda$, but independent of the 
zero-momentum gauge field. This term will be dropped, as its value 
and its derivative with respect to $\lambda$, at $\lambda=0$, can be 
interpreted as the perturbative contribution to the vacuum energy 
$E_0$, resp. $\langle0|\norm{A}_{\rm min}^2|0\rangle$. Keeping only 
the terms in the effective hamiltonian that depend on the 
zero-momentum gauge field (and on $\lambda$) we may calculate the 
groundstate energy as a function of $\lambda$. Its derivative at 
$\lambda=0$ gives the finite volume non-perturbative contribution 
to $\langle0|\norm{A}_{\rm min}^2|0\rangle$. 

\begin{figure}[htb]
\vspace{5cm}
\includegraphics{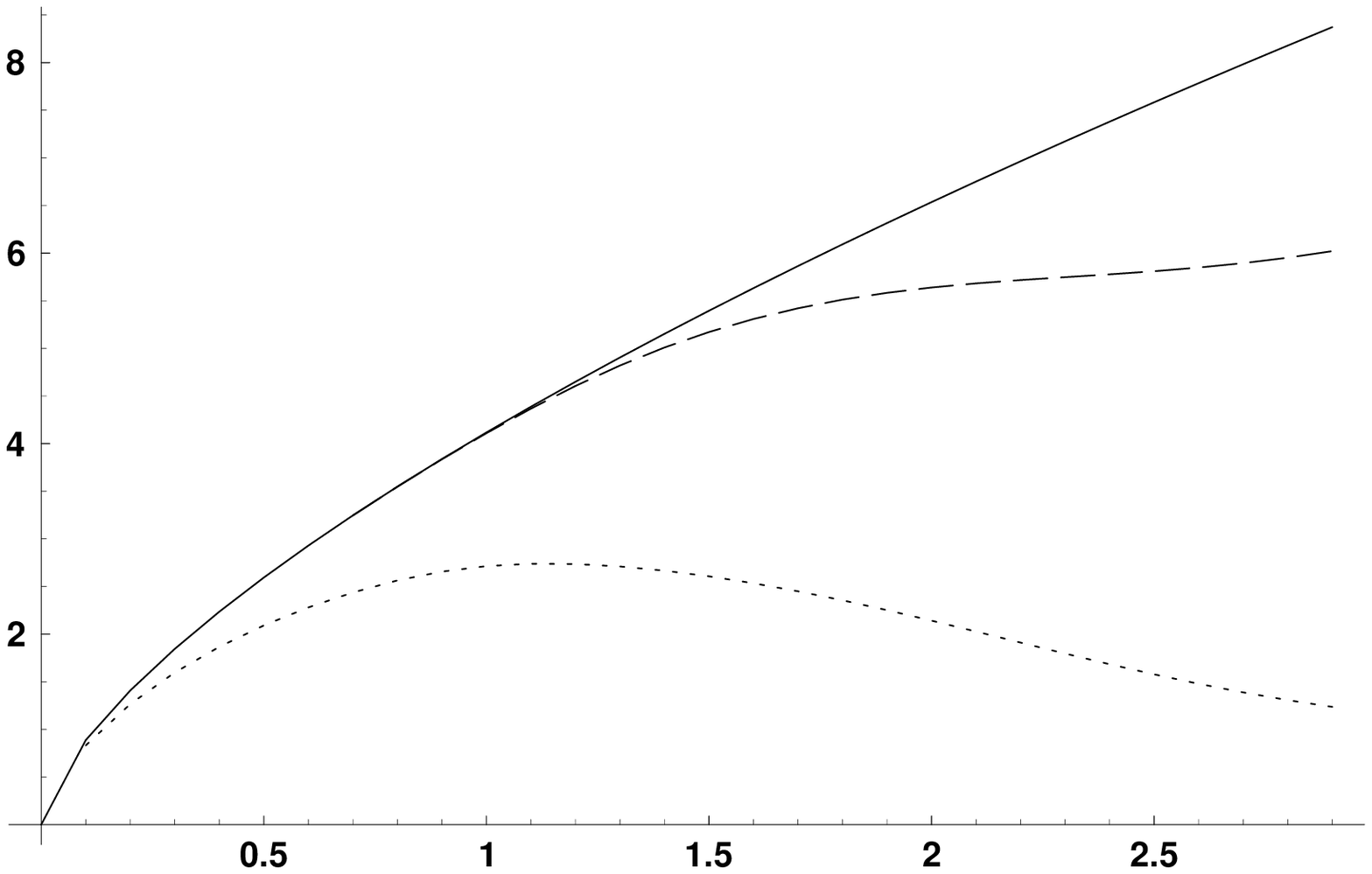}
\includegraphics{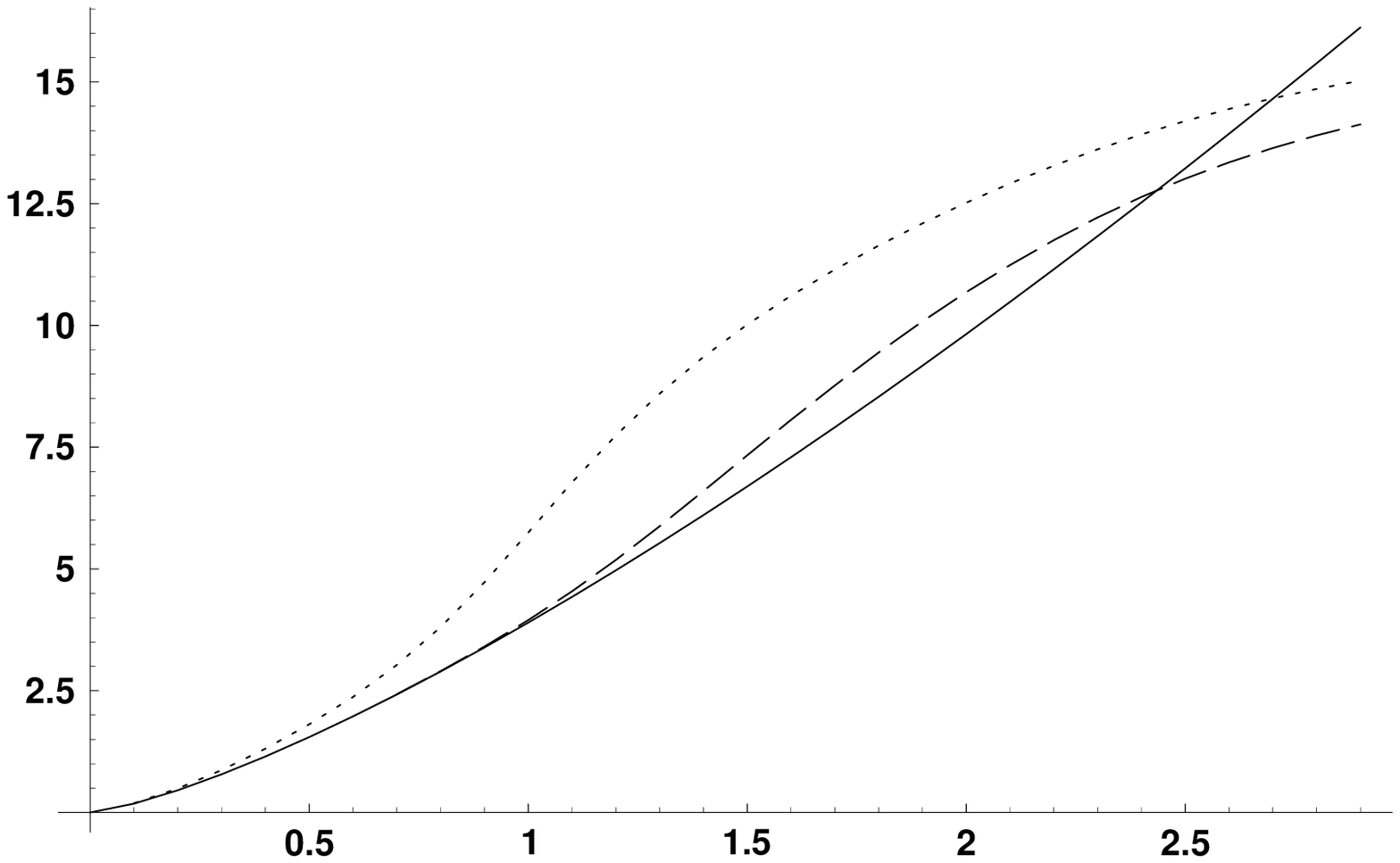}
\caption{On the left is shown $LE_0$ and on the right $L^2\langle
A^2\rangle$ for $SU(2)$, after subtracting the perturbative contribution 
as discussed in the text, both as a function of $g(L)$. The full 
lines give the lowest order result, $4.11672 g^{2/3}(L)$, resp. 
$3.89775 g^{4/3}(L)$. The dashed lines include the effect of the 
boundary conditions, for the lowest order effective hamiltonian. 
The dotted lines include higher order contributions to the effective 
hamiltonian (see the discussion in the text).}\label{fig:A2}
\end{figure}

Adding a mass term in the zero-momentum sector seems to remove the 
quartic nature of the potential. However, as rescaling the 
zero-momentum component of the gauge field with $g^{2/3}$ reveals, 
the term quadratic in the zero-momentum gauge field is proportional 
to $g^{4/3}$, and hence of lower order. As a matter of fact, the 
quantum corrections induce a term of this order in the effective 
hamiltonian~\cite{Lue}. In Fig.~\ref{fig:A2} we illustrate for $SU(2)$ 
the lowest order result, with and without incorporating the boundary 
conditions, based on Ref.~\cite{Koll} (in terms of the terminology 
introduced there, the lowest order result is type IIIA with, and 
type IIIC without incorporating the boundary conditions; the computer 
code used here is essentially the one developed for that paper). 
In this figure we have shown the result up to $g=2.9$, which 
corresponds roughly to a volume of a cubic fermi. For larger volumes 
the wave functional will have spread in other directions as well, such 
that the perturbative approximation for these non-zero momentum modes 
no longer holds. Here we only wish to illustrate the influence boundary 
conditions in field space can have. It should, however, be understood 
that where a deviation starts to occur, the coupling is already sizable 
and a fully self-consistent calculation requires us to also include 
the higher order contributions to the effective hamiltonian. To give a 
flavor of the magnitude of these corrections we show with the dotted 
curves the result obtained when ignoring the dependence of the 
groundstate wave function on $\lambda$.

\section{Conclusion}

We have shown how we can in principle define $\langle A^2\rangle$
as $L^{-3}\langle0|\norm{A}_{\rm min}^2|0\rangle$ in a hamiltonian 
formalism, which restricts $A$ to the fundamental domain. No two 
gauge fields in the interior of the domain are gauge equivalent, 
but gauge fields on the boundary in general do have gauge copies, 
also on the boundary. Crucial for defining 
$\langle0|\norm{A}_{\rm min}^2|0\rangle$ is that $\norm{A}_{\rm min}$ 
takes on the same value for these gauge equivalent gauge fields (as is 
intrinsic to the definition of the boundary of the fundamental domain).

To lowest order, in a small volume, this expectation value is generated
in the usual way by adding the appropriate source to the hamiltonian.
The resulting effective hamiltonian is specified in terms of the gauge 
field components that will feel the boundary of the fundamental domain, 
which in a cubic finite volume smaller than about 0.75 cubic fermi 
means an effective hamiltonian in terms of the zero-momentum modes. 
Although in this domain there is near perfect agreement with the 
low-lying spectrum obtained from lattice gauge theory, it is known 
that larger volumes are required to get close to the infinite volume 
limit. Nevertheless, we have shown a framework in which we can make 
sense of this quantity, even though its applicability is for 
technical reasons still limited to a finite volume.

\section*{Acknowledgements}

PvB is grateful for the hospitality extended to him at the MPI in
Munich.

\end{document}